\documentclass{article}
\usepackage{amsmath, graphicx, hyperref, natbib}
\usepackage{amssymb}
\usepackage{algorithm}
\usepackage{algpseudocode}
\usepackage{xcolor}
\usepackage[margin=1in]{geometry}

\title{Data-Adaptive Identification of Effect Modifiers through Stochastic Shift Interventions and Cross-Validated Targeted Learning}

\author{David McCoy, Wenxin Zhang, 
Alan Hubbard, \\ Mark van der Laan, Alejandro Schuler  \\ \hspace{.5cm}\\
    Division of Biostatistics, University of California, Berkeley}

\begin{document}

\maketitle

\begin{abstract}
{In epidemiology, identifying subpopulations that are particularly vulnerable to exposures and those who may benefit differently from exposure-reducing interventions is essential. Factors such as age, gender-specific vulnerabilities, and physiological states such as pregnancy are critical for policymakers when setting regulatory guidelines. However, current semi-parametric methods for estimating heterogeneous treatment effects are often limited to binary exposures and can function as black boxes, lacking clear, interpretable rules for subpopulation-specific policy interventions. This study introduces a novel method that uses cross-validated targeted minimum loss-based estimation (TMLE) paired with a data-adaptive target parameter strategy to identify subpopulations with the most significant differential impact of simulated policy interventions that reduce exposure. Our approach is assumption-lean, allowing for the integration of machine learning while still yielding valid confidence intervals. We demonstrate the robustness of our methodology through simulations and application to data from the National Health and Nutrition Examination Survey. Our analysis of NHANES data on persistent organic pollutants (POPs) and leukocyte telomere length (LTL) identified age as a significant effect modifier. Specifically, we found that exposure to 3,3',4,4',5-pentachlorobiphenyl (PCNB) consistently had a differential impact on LTL, with a one-standard deviation reduction in exposure leading to a more pronounced increase in LTL among younger populations compared to older ones. We offer our method as an open-source software package, \texttt{EffectXshift}, enabling researchers to investigate the effect modification of continuous exposures. The \texttt{EffectXshift} package provides clear and interpretable results, informing targeted public health interventions and policy decisions.}
\end{abstract}

\section{Introduction}

Environmental risk factors—including pollutants and social determinants—pose significant health risks, but their impact is not uniformly distributed across populations. Certain subpopulations—such as children, pregnant women, the elderly, and socioeconomically disadvantaged communities—are disproportionately exposed to harmful conditions, leading to pronounced health disparities \cite{balbus_identifying_2009, ruiz_disparities_2018, zorzetto_confounder-dependent_2024} For instance, studies have shown that African Americans, Latinos, and low-income individuals are more likely to experience higher exposures to environmental pollutants like air pollution and endocrine-disrupting compounds (EDCs) such as polychlorinated biphenyls (PCBs) and bisphenol A (BPA), as well as adverse social conditions, all of which are linked to metabolic disorders and diabetes \cite{ruiz_disparities_2018}. Factors contributing to these disparities include dietary patterns, use of certain consumer products, residential proximity to pollution sources, and social determinants like socioeconomic status and access to healthcare \cite{MAKRI2008326}.

Identifying these vulnerable subpopulations is crucial for designing targeted public health interventions and policies aimed at reducing exposure and mitigating health risks. \emph{Effect modification}—the phenomenon where the effect of an exposure on an outcome differs across levels of another variable - is a key concept in epidemiological research for uncovering such differential impacts \cite{Rothman2008}. Traditional methods for assessing effect modification, such as stratified analyses, interaction terms in regression models, and generalized additive models (GAMs), have been widely used \cite{Samoli2011, Analitis2014}. For example, Samoli et al. \cite{Samoli2011} examined how air pollution affects pediatric asthma exacerbations by stratifying analyses by age, sex, and season. However, these conventional methods have limitations when dealing with complex environmental exposures, which are often continuous, multidimensional, and may act synergistically as mixtures \cite{gibson2019}.

While Bayesian methods such as Bayesian profile regression and confounder-dependent mixture modeling have shown promise in identifying subpopulations susceptible to multipollutant exposures \cite{coker_multi-pollutant_2018, Zorzetto2024}, they can be computationally intensive for very large datasets and may require specialized expertise to implement and interpret. Similarly, meta-learning methods for estimating heterogeneous treatment effects, like the T-learner, S-learner, and X-learner \cite{kunzel2019metalearners}, while powerful, often focus on binary treatments and may lack the interpretability desired in environmental studies dealing with continuous exposures. Our approach aims to address these challenges by providing a computationally efficient, interpretable method that naturally handles continuous exposures and leverages machine learning techniques within a causal inference framework.

To address these challenges, we propose a novel, data-adaptive method that leverages causal inference techniques with stochastic shift interventions to identify subpopulations most vulnerable to harmful exposures. \emph{Stochastic shift interventions} modify the distribution of exposures rather than setting them to fixed values, making them particularly relevant for environmental policies aiming to reduce pollutant levels across populations \cite{DiazMunoz2012, Kennedy2019}. By integrating targeted maximum likelihood estimation (TMLE) with machine learning, our approach allows for flexible modeling of complex exposure-outcome relationships while providing valid statistical inference \cite{vanderLaan2011}.

To illustrate the practical significance of our method, consider the exposure to per- and polyfluoroalkyl substances (PFAS), a group of pollutants found in various consumer products and associated with adverse health outcomes such as thyroid cancer \cite{Jain2016}. Suppose that we have biomarker data on PFAS levels from blood samples, thyroid cancer outcomes, and baseline covariates such as age, body mass index (BMI), use of consumer products, and other relevant factors. Our study aims to inform analysts and policymakers about which pollutants in the PFAS mixture has the largest differential impact if reduced in certain subpopulations. For example, we might find that reducing exposure to perfluorononanoic acid (PFNA) by 5 parts per billion (ppb) leads to a 12\% decrease in thyroid-related diseases among pregnant women aged 18--29, compared to a 3\% decrease in other individuals. Such insights enable the development of targeted interventions and policies that focus on the most vulnerable groups and the most impactful exposures.

Our method achieves two key objectives: (i) identifying the exposure-covariate relationships where an intervention would have the maximal differential impact, and (ii) estimating the effects of such interventions within the identified subpopulations using an assumption-lean estimator. This approach enhances interpretability by providing clear rules for subpopulation-specific interventions and addresses limitations of existing methods that function as black boxes.

This paper is structured as follows. Section \ref{sec:stochastic_shift_interventions} introduces stochastic shift interventions, providing the foundational concepts and notation. In Section \ref{sec:oracle_target_parameter}, we discuss our oracle target parameter, which defines the maximal effect modifier relationship. Section \ref{sec:algorithm} describes the recursive partitioning algorithm designed to estimate this oracle parameter, identifying subregions within the covariates that exhibit the maximal differential impact of a stochastic intervention. Section \ref{sec:simulation_study} presents simulations of a mixed exposure scenario, and \ref{sec:simulation_results} shows the identification and accurate estimation of the maximal effect modifier when applying our method to simulatino data. In Section \ref{sec:nhanes_application}, we apply our approach to NHANES data to explore the effects of mixed POPs on telomere length, assessing possible effect modification. We close with a conclusion in \ref{sec:conclusion}. The code for our method, along with the applications and simulations, is available on GitHub in our open-source package called \href{https://github.com/blind-contours/EffectXshift}{\texttt{EffectXshift}}.

\section{Defining Stochastic Shift Interventions}
\label{sec:stochastic_shift_interventions}

Consider an experiment where an exposure variable \( A \in \mathcal{A} \), an outcome \( Y \), and a set of covariates \( W \in \mathcal{W} \) are measured for \( n \) randomly sampled units, with \( O = (W, A, Y) \) denoting a random variable drawn from a distribution \( P_0 \). Let \( O_1, \ldots, O_n \) be \( n \) i.i.d. observations of \( O \).

We posit a Nonparametric Structural Equation Model (NPSEM) that encodes the data-generating process:
\[
W = f_W(U_W), \quad A = f_A(W,U_A), \quad Y = f_Y(W,A,U_Y),
\]
where \( U_W, U_A, U_Y \) are exogenous random variables and \( f_W, f_A, f_Y \) are deterministic functions. Under this model, the observed data distribution factors as:
\[
P_0(O) = P_{0,W}(W) \, P_{0,A}(A \mid W) \, P_{0,Y}(Y \mid A,W).
\]

Define a function 
\[
g_0: \mathcal{A} \times \mathcal{W} \to \mathbb{R}^+,
\]
and use \( g_0(a \mid w) \) as shorthand for the conditional density of \( A \) given \( W = w \). Let \(\overline{Q}_0(a,w) = E_0[Y \mid A = a, W = w]\) and let \( p_{0,W}(w) \) denote the density of \( W \).

We now consider a \emph{stochastic shift intervention} that modifies the exposure mechanism. Instead of drawing \( A \) from \( g_0(a \mid w) \) as originally defined, we define a shifted exposure distribution:
\[
p_{0,A_{\delta}}(a \mid w) = g_0(a-\delta \mid w),
\]
where \(\delta \in \mathbb{R}\) characterizes the magnitude and direction of the shift. Under this modified NPSEM, each unit's exposure is effectively shifted by \(\delta\) units, and the corresponding counterfactual outcome \( Y_{A_\delta} \) is the outcome that would have been observed if the unit's exposure had been drawn from the shifted distribution \( p_{0,A_{\delta}} \).

Our parameter of interest is:
\[
\Psi(P_0) = E_{P_{0,A_{\delta}}}[Y_{A_{\delta}}],
\]
the expected outcome under the stochastic shift intervention.

Identification of \(\Psi(P_0)\) from observed data relies on standard causal conditions:

\paragraph{Exchangeability (No Unmeasured Confounding)} Under exchangeability, \( Y_a \perp A \mid W \) for all \( a \in \mathcal{A} \). Given \( W \), the observed assignment of \( A \) can be viewed as if it were randomized. This assumption ensures that the conditional associations we observe can be interpreted as reflecting causal effects of the exposure on the outcome.

\paragraph{Positivity} The positivity condition requires that the shift \(\delta\) is chosen so that the shifted exposure levels remain within the support of \( A \) for each covariate stratum. Let \( l(w) \) and \( u(w) \) be the minimum and maximum values of \( A \) attainable for units with \( W = w \). To maintain positivity, we must ensure that for all \( (a,w) \) in the support of \((A,W)\), the point \( a-\delta \) also lies within the interval \([l(w), u(w)]\). If \( a-\delta \) falls outside this support, then \( g_0(a-\delta \mid w) = 0 \) for those values, violating positivity and preventing identification of \(\Psi(P_0)\).

In practice, this means selecting \(\delta\) so that the shifted exposure remains in the feasible range defined by \( l(w) \) and \( u(w) \). When this is satisfied, there exists an \(\epsilon > 0\) such that
\[
0 < \frac{g_0(a-\delta \mid w)}{g_0(a \mid w)} \leq \frac{1}{\epsilon}
\]
for all relevant \((a,w)\), ensuring that the stochastic shift intervention is well-defined and that the resulting counterfactual distribution \( P_{0,A_\delta} \) is identifiable from the observed data.

\paragraph{Interpretation} Intuitively, \(\Psi(P_0)\) represents the expected outcome if we replaced the original exposure distribution \( g_0(a \mid w) \) with one shifted by \(\delta\). This parameter captures the causal effect of a hypothetical population-level intervention that systematically reduces exposure levels. By estimating \(\Psi(P_0)\), we can assess the global impact of such an intervention and compare effects across subpopulations defined by \( W \), thereby guiding targeted interventions and informing policy decisions. Because the estimation relies only on observational data and these well-defined causal assumptions (exchangeability and positivity), it allows us to infer the potential effects of interventions that have not been experimentally evaluated in practice.

\subsection{Efficient Estimation and Inference for Stochastic Interventions}

To estimate the causal effect of stochastic interventions effectively, we aim to construct a semiparametric efficient estimator. Such an estimator achieves the smallest possible asymptotic variance among regular asymptotically linear (RAL) estimators, satisfying:
\begin{equation}
    \sqrt{n} (\hat{\Psi} - \Psi) = \sqrt{n}\mathbb{E}_n[\phi(O)] + o_p(1),
\end{equation}
where \( \phi(O) \) is the efficient influence function, and \( \mathbb{E}_n \) denotes the empirical mean. By the Central Limit Theorem, RAL estimators achieve the asymptotic variance bound \( \mathbb{E}[\phi^2(O)] \), facilitating valid statistical inference.

In the context of our stochastic shift intervention, the efficient influence function \( \phi_P(O) \) of \( \Psi \) at distribution \( P \) is given by:
\begin{equation}
\phi_P(O) = \frac{g_0(A - \delta \mid W)}{g_0(A \mid W)} \left( Y - \overline{Q}_0(A, W) \right) + \overline{Q}_0(A - \delta, W) - \Psi(P).
\end{equation}

Several estimators can be employed for estimating our stochastic shift parameter, including inverse probability weighting (IPW), augmented inverse probability weighting (AIPW), and targeted maximum likelihood estimation (TMLE). However, IPW estimators can be inefficient and sensitive to model misspecification, and AIPW estimators may not respect the bounds of the outcome, potentially resulting in greater variability. Therefore, we choose TMLE to estimate our parameter due to its efficiency and robustness in finite samples \cite{Li2022}.

Targeted learning begins with initial estimates of the outcome regression \( \hat{\overline{Q}}(A, W) \) and the exposure density \( \hat{g}(A \mid W) \), which can be obtained through flexible machine learning techniques. These initial estimates are then updated using a clever covariate based on the efficient influence function. The refinement process involves fitting a parametric fluctuation model that targets the parameter of interest. This results in a final estimate \( \hat{\overline{Q}}^\star(A - \delta, W) \) that is efficient and less biased. The updated estimate is then used to compute the TMLE of \( \Psi \), and the influence function is utilized to derive confidence intervals.

\section{Causal Subpopulation Intervention Effects}
\label{sec:oracle_target_parameter}

Our primary objective in this paper is not only to estimate the overall effects of environmental exposures but also to discern how these effects differ across various subpopulations. Within the framework of stochastic shift interventions, our focus is on identifying specific types of individuals from baseline characteristics that change the impact of policy interventions, such as reducing exposure to pollutants. Unlike the previously described stochastic shift parameter, which averages effects across the entire population, we aim to define our causal parameter for particular subgroups. This allows us to pinpoint the subpopulations where the intervention yields the maximum differential effect, thereby providing more targeted and effective public health strategies.

\subsection{Oracle Parameter for Maximal Difference in Expected Outcomes in Subregions}

The oracle parameter, denoted as \(\psi^*\), represents the maximal differential impact of an intervention in a specific subregion of the covariate space. It is defined as follows:

\begin{align}
\psi^* &= \underset{A_i \in \boldsymbol{A}, \, \mathcal{V} \subseteq \mathcal{W}}{\sup} 
\left\{ \mathbb{E}\left[ Y_{A_i - \delta_i, \mathbf{A}_{-i}} - Y \mid W \in \mathcal{V} \right] \right.  \nonumber \\
&\left. \quad - \mathbb{E}\left[ Y_{A_i - \delta_i, \mathbf{A}_{-i}} - Y \mid W \in \mathcal{V}^c \right]
\right\},
\end{align}

where:
\begin{itemize}
    \item \(\boldsymbol{A} = (A_1, A_2, \ldots, A_p)\) denotes the vector of exposures in the mixture, with \(A_i\) representing the \(i\)-th exposure and \(\mathbf{A}_{-i}\) denoting all exposures excluding \(A_i\).
    \item \(\mathcal{V}\) and \(\mathcal{V}^c\) denote a subregion of the covariate space \(\mathcal{W}\) and its complement, respectively.
    \item \(Y_{A_i - \delta_i, \mathbf{A}_{-i}}\) represents the potential outcome under a stochastic shift intervention where exposure \(A_i\) is shifted by \(\delta_i\), while other exposures \(\mathbf{A}_{-i}\) remain unchanged.
\end{itemize}

\(\psi^*\) identifies the combination of exposure \(A_i\) and subregion \(\mathcal{V}\) that maximizes the difference in expected outcomes between \(\mathcal{V}\) and its complement \(\mathcal{V}^c\). This parameter captures effect modification, highlighting where the intervention has the most substantial and relevant impact.

In general, we are interested in the causal exposure-covariate region combination where the effect of the intervention, compared to the complementary covariate space, is maximally different, controlling for other exposures. From now on, we refer to these regions as \(\mathcal{V}\) and \(\mathcal{V}^c\).

This method allows us to identify and focus on subpopulations where the intervention has the most substantial and relevant impact, thereby informing targeted public health strategies and interventions.

\subsection{Statistical Subpopulation Intervention Effects}

This section reformulates the causal parameter of interest in terms of observable quantities, thereby defining a statistical parameter that can be estimated from the data. We consider a fixed subpopulation \(\mathcal{V} \subseteq \mathcal{W}\), but emphasize that subpopulation-specific parameters are interpreted as features of the full population distribution. Let \(\pi_{\mathcal{V}} = P(W \in \mathcal{V})\) denote the probability that a randomly selected unit from the full population belongs to \(\mathcal{V}\).

Under standard identification assumptions, including no unmeasured confounding and positivity, the causal parameter \(\psi^*\) is identified as:
\begin{align}
\psi^* &= \sup_{A_i \in \boldsymbol{A},\; \mathcal{V} \subseteq \mathcal{W}} 
\left\{ \mathbb{E}\bigl[ \overline{Q}_0(A_i - \delta_i, \mathbf{A}_{-i}, W) - \overline{Q}_0(A_i, \mathbf{A}_{-i}, W) \mid W \in \mathcal{V} \bigr] \right. \nonumber \\
&\left. \quad - \mathbb{E}\bigl[ \overline{Q}_0(A_i - \delta_i, \mathbf{A}_{-i}, W) - \overline{Q}_0(A_i, \mathbf{A}_{-i}, W) \mid W \in \mathcal{V}^c \bigr] \right\},
\end{align}
where \(\overline{Q}_0(a_i, \mathbf{a}_{-i}, w) = \mathbb{E}[Y \mid A_i=a_i, \mathbf{A}_{-i}=\mathbf{a}_{-i}, W=w]\).

Fixing \(\mathcal{V}\), we define the statistical parameter \(\psi_{\mathcal{V}}\) as:
\begin{align}
\psi_{\mathcal{V}} &= \max_{A_i \in \boldsymbol{A}} \bigl( \Psi_{\mathcal{V}}(P) - \Psi_{\mathcal{V}^c}(P) \bigr),
\end{align}
where
\begin{align}
\Psi_{\mathcal{V}}(P) &= \frac{\mathbb{E}[\Delta_i(W)\mathbb{I}\{W \in \mathcal{V}\}]}{\pi_{\mathcal{V}}}, \quad
\Psi_{\mathcal{V}^c}(P) = \frac{\mathbb{E}[\Delta_i(W)\mathbb{I}\{W \in \mathcal{V}^c\}]}{1-\pi_{\mathcal{V}}}, \\
\Delta_i(W) &= \overline{Q}_0(A_i - \delta_i, \mathbf{A}_{-i}, W) - \overline{Q}_0(A_i, \mathbf{A}_{-i}, W).
\end{align}

This formulation ensures that \(\Psi_{\mathcal{V}}(P)\) and \(\Psi_{\mathcal{V}^c}(P)\) are understood as expectations taken over the full population distribution, normalized by \(\pi_{\mathcal{V}}\) and \(1-\pi_{\mathcal{V}}\) respectively. The parameter \(\psi_{\mathcal{V}}\) thus represents a feature of the entire population, focusing attention on differences between a particular subpopulation \(\mathcal{V}\) and its complement.

\paragraph{Influence Functions Corresponding to the Parameters}

To perform inference on \(\Psi_{\mathcal{V}}(P)\) and \(\Psi_{\mathcal{V}^c}(P)\), we rely on their efficient influence functions (EIFs), which are essential for constructing asymptotically efficient estimators and valid confidence intervals. The EIFs incorporate the normalization by \(\pi_{\mathcal{V}}\):

\begin{align}
\phi_{\mathcal{V}}(O) &= \frac{\mathbb{I}\{W \in \mathcal{V}\}}{\pi_{\mathcal{V}}} \left[ \frac{g_0(A_i - \delta_i \mid W)}{g_0(A_i \mid W)} \bigl(Y - \overline{Q}_0(A_i,\mathbf{A}_{-i},W)\bigr) + \Delta_i(W) - \Psi_{\mathcal{V}}(P) \right] \nonumber \\
&\quad + \bigl(\mathbb{I}\{W \in \mathcal{V}\} - \pi_{\mathcal{V}}\bigr)\frac{\Psi_{\mathcal{V}}(P)}{\pi_{\mathcal{V}}}, \\
\phi_{\mathcal{V}^c}(O) &= \frac{\mathbb{I}\{W \in \mathcal{V}^c\}}{1-\pi_{\mathcal{V}}} \left[ \frac{g_0(A_i - \delta_i \mid W)}{g_0(A_i \mid W)} \bigl(Y - \overline{Q}_0(A_i,\mathbf{A}_{-i},W)\bigr) + \Delta_i(W) - \Psi_{\mathcal{V}^c}(P) \right] \nonumber \\
&\quad + \bigl(\mathbb{I}\{W \in \mathcal{V}^c\} - (1-\pi_{\mathcal{V}})\bigr)\frac{\Psi_{\mathcal{V}^c}(P)}{1-\pi_{\mathcal{V}}}.
\end{align}

\paragraph{Estimation and Inference}

In practice, the true functions \(\overline{Q}_0\) and \(g_0\), as well as \(\pi_{\mathcal{V}}\), are unknown and must be estimated from observed data. Let \(\hat{\pi}_{\mathcal{V}} = \frac{1}{n}\sum_{i=1}^n \mathbb{I}\{W_i \in \mathcal{V}\}\). We define \(\hat{g}(A \mid W)\) as a suitable estimator of \(g_0(A \mid W)\), the conditional density of \(A\) given \(W\). The function \(\overline{Q}_0(A,W)\) is estimated by \(\hat{\overline{Q}}(A,W)\), and we employ Targeted Maximum Likelihood Estimation (TMLE) to refine this initial estimate, resulting in \(\hat{\overline{Q}}^\ast(A,W)\). The superscript \(\ast\) indicates that the estimate has been updated in a manner that targets the parameter of interest, thereby reducing bias and improving efficiency.

Once \(\hat{\overline{Q}}^\ast\) and \(\hat{g}\) are obtained, we define:
\begin{align}
\hat{\Delta}_i(W_i) &= \hat{\overline{Q}}^\ast(A_i - \delta_i, \mathbf{A}_{-i},W_i) - \hat{\overline{Q}}^\ast(A_i,\mathbf{A}_{-i},W_i).
\end{align}

We then estimate \(\Psi_{\mathcal{V}}(P)\) and \(\Psi_{\mathcal{V}^c}(P)\) as:
\begin{align}
\hat{\Psi}_{\mathcal{V}} &= \frac{\sum_{i=1}^n \hat{\Delta}_i(W_i)\mathbb{I}\{W_i \in \mathcal{V}\}}{\hat{\pi}_{\mathcal{V}}}, \quad
\hat{\Psi}_{\mathcal{V}^c} = \frac{\sum_{i=1}^n \hat{\Delta}_i(W_i)\mathbb{I}\{W_i \in \mathcal{V}^c\}}{1-\hat{\pi}_{\mathcal{V}}}.
\end{align}

Let \(n_{\mathcal{V}} = \sum_{i=1}^n \mathbb{I}\{W_i \in \mathcal{V}\}\) and \(n_{\mathcal{V}^c} = \sum_{i=1}^n \mathbb{I}\{W_i \in \mathcal{V}^c\}\). The parameter of interest, representing the difference in subpopulation-specific effects, is estimated by:
\[
\hat{\psi}_{\mathcal{V}} = \hat{\Psi}_{\mathcal{V}} - \hat{\Psi}_{\mathcal{V}^c}.
\]

\paragraph{Variance Estimation and Statistical Inference}

To derive standard errors and confidence intervals, we use the estimated influence functions:
\[
\hat{\phi}_{\mathcal{V}}(O_i), \quad \hat{\phi}_{\mathcal{V}^c}(O_i),
\]
obtained by plugging the estimated quantities \(\hat{\overline{Q}}^\ast\), \(\hat{g}\), and \(\hat{\pi}_{\mathcal{V}}\) into the EIF formulas.

The variance of \(\hat{\psi}_{\mathcal{V}}\) is:
\[
\widehat{\mathrm{Var}}(\hat{\psi}_{\mathcal{V}}) = \widehat{\mathrm{Var}}(\hat{\Psi}_{\mathcal{V}}) + \widehat{\mathrm{Var}}(\hat{\Psi}_{\mathcal{V}^c}),
\]
assuming negligible covariance due to disjoint subpopulations. The variance estimators require incorporating the appropriate scaling by \(1/n_{\mathcal{V}}\) and \(1/n_{\mathcal{V}^c}\). After computing these variances from the EIFs, a normal approximation can be employed to construct confidence intervals and conduct hypothesis tests, such as testing whether \(\psi_{\mathcal{V}}=0\).

In summary, defining the parameter \(\psi_{\mathcal{V}}\) as a population-level feature involving a particular subpopulation \(\mathcal{V}\) and using the normalized EIFs ensures valid statistical inference. By applying TMLE and variance estimation through the EIFs, this framework provides efficient, assumption-lean estimates for subpopulation-specific intervention effects and the differences between subpopulations. Such estimates facilitate informed decision-making and targeted interventions in settings where heterogeneous effects are of interest.

\paragraph{Incorporating Variability in Region Selection}

Although the statistical parameter \(\psi_{\mathcal{V}}\) quantifies the difference in expected outcomes between the subpopulation \(\mathcal{V}\) and its complement \(\mathcal{V}^c\), it is essential to consider the variability of these estimates. Even a large observed difference may not be statistically significant once variability is taken into account. By leveraging the influence functions and their corresponding variance estimates, we can formally assess the statistical significance of the observed effect modification.

To test the null hypothesis \( H_0: \psi_{\mathcal{V}} = 0 \), we construct a t-statistic:
\[
T_{\mathcal{V}} = \frac{\hat{\psi}_{\mathcal{V}}}{\sqrt{\widehat{\mathrm{Var}}(\hat{\psi}_{\mathcal{V}})}}.
\]

Under standard regularity conditions, \( T_{\mathcal{V}} \) converges in distribution to a standard normal random variable. Consequently, we can employ conventional inferential procedures (e.g., using standard normal quantiles) to determine whether the observed difference in intervention effects is statistically significant. If \( |T_{\mathcal{V}}| \) exceeds a chosen critical value (e.g., corresponding to a two-sided \(\alpha = 0.05\) level), we reject \( H_0 \) and conclude that there is evidence of a meaningful difference in the intervention effects between \(\mathcal{V}\) and \(\mathcal{V}^c\).

\paragraph{Remark}

In applied settings, it is crucial to consider both the magnitude and statistical significance of the estimated effect difference. Focusing solely on the point estimate without accounting for variability risks identifying subpopulations based on random fluctuations rather than true differences. By incorporating a careful variance-based assessment, researchers can more confidently identify meaningful subpopulations for targeted interventions and informed policy decisions.

\subsection{Identifying Subpopulations with Maximal Effect Difference}
\label{sec:algorithm}

Building upon the statistical parameters and influence functions defined in the previous section, we now describe a procedure for identifying subpopulations with maximal effect modification for a given exposure. The estimation involves a two-stage approach:

\textbf{First Stage:} We obtain efficient estimates of the Individual Intervention Effects (IIEs) and the corresponding Influence Curve Estimates (ICEs) for each exposure \( A_k \) in the mixture \( \boldsymbol{A} = (A_1, A_2, \ldots, A_p) \).

\textbf{Second Stage:} Utilizing these estimates, we apply a greedy partitioning algorithm to find covariate regions \( \mathcal{V} \) where the effect modification \(\psi_{\mathcal{V}}\) is maximized and statistically significant.

\subsubsection*{First Stage: Estimating IIEs and ICEs}

For each exposure \( A_k \), we proceed as follows:

\begin{enumerate}
    \item \textbf{Initial Estimation of Outcome Regression:}  
    Fit an initial estimator \(\hat{\overline{Q}}(A, W)\) for \(\mathbb{E}[Y \mid \boldsymbol{A}, W]\) using flexible machine learning methods (e.g., Super Learner).

    \item \textbf{Estimating the Density Ratio:}  
    Estimate \(\hat{g}(A_k \mid W)\) and \(\hat{g}(A_k - \delta_k \mid W)\), or directly estimate the density ratio \(\hat{r}(A_k, W) = \frac{\hat{g}(A_k - \delta_k \mid W)}{\hat{g}(A_k \mid W)}\).

    \item \textbf{Debiasing with TMLE:}  
    Update the initial estimator \(\hat{\overline{Q}}(A, W)\) to \(\hat{\overline{Q}}^\ast(A, W)\) via Targeted Maximum Likelihood Estimation (TMLE), targeting the parameter \(\Psi_{\mathcal{V}}(P)\) for the chosen exposure. This involves:
    \begin{enumerate}
        \item Constructing the clever covariate \(H(A_k, W) = \hat{r}(A_k, W)\).
        \item Fitting a parametric model (e.g., logistic regression for binary outcomes) of \(Y\) on \(H(A_k, W)\) with offset \(\text{logit}(\hat{\overline{Q}}(A_k, W))\) for binary \(Y\), or a suitable quasi-likelihood model for continuous \(Y\).
        \item Obtaining \(\hat{\overline{Q}}^\ast(A, W)\), the updated estimator incorporating the TMLE fluctuation.
    \end{enumerate}

    \item \textbf{Calculating Individual-Level Estimates:} For each individual \( i \):
    \begin{enumerate}
        \item \textbf{Individual Intervention Effect (IIE):}  
        Compute the estimated IIE for exposure \( A_k \):
        \begin{equation}
        \hat{\Delta}_{ik}(W_i) = \hat{\overline{Q}}^\ast(A_k - \delta_k,\, \mathbf{A}_{-k},\, W_i) - \hat{\overline{Q}}^\ast(A_k,\, \mathbf{A}_{-k},\, W_i).
        \end{equation}

        \item \textbf{Influence Curve Estimate (ICE):}  
        Compute the estimated influence function:
        \begin{equation}
        \hat{\phi}_{ik}(O_i) = \frac{\hat{g}(A_k - \delta_k \mid W_i)}{\hat{g}(A_k \mid W_i)} \bigl(Y_i - \hat{\overline{Q}}^\ast(A_k, W_i)\bigr) + \hat{\Delta}_{ik}(W_i) - \hat{\Psi}_k,
        \end{equation}
        where \(\hat{\Psi}_k\) is the estimated population-level parameter (e.g., the average intervention effect for exposure \( A_k \) in the sample).
    \end{enumerate}
\end{enumerate}

By iterating through each exposure \( A_k \), we construct matrices of IIEs \(\hat{\Delta}\) and ICEs \(\hat{\phi}\), where rows correspond to individuals \( i \) and columns correspond to exposures \( k \).

\subsubsection*{Second Stage: Identifying Covariate Regions with Maximal Effect Modification}

In the second stage, we utilize the estimated IIEs \(\hat{\Delta}_{ik}\) and ICEs \(\hat{\phi}_{ik}\) for a chosen exposure \( A_k \) to identify covariate regions \(\mathcal{V}\) where the effect modification \(\psi_{\mathcal{V}}\) is maximized. We employ a greedy partitioning algorithm based on statistical significance, ensuring that the identified regions exhibit both substantial and statistically significant differences.

The algorithm proceeds as follows:

\begin{enumerate}
    \item \textbf{Initialization:}  
    Start with the full covariate space \(\mathcal{W}\).

    \item \textbf{Greedy Search for Optimal Split:}  
    For each covariate \( W_j \) and each candidate split point \( s \):
    \begin{enumerate}
        \item Define \(\mathcal{V} = \{ i : W_{ij} \leq s \}\) and \(\mathcal{V}^c = \{ i : W_{ij} > s \}\).
        \item Compute the estimated effect modification:
        \[
        \hat{\psi}_{\mathcal{V}} = \hat{\Psi}_{\mathcal{V}} - \hat{\Psi}_{\mathcal{V}^c},
        \]
        where
        \[
        \hat{\Psi}_{\mathcal{V}} = \frac{1}{n_{\mathcal{V}}}\sum_{i \in \mathcal{V}}\hat{\Delta}_{ik}(W_i), \quad
        \hat{\Psi}_{\mathcal{V}^c} = \frac{1}{n_{\mathcal{V}^c}}\sum_{i \in \mathcal{V}^c}\hat{\Delta}_{ik}(W_i).
        \]

        \item Estimate the variance \(\widehat{\mathrm{Var}}(\hat{\psi}_{\mathcal{V}})\) using the ICEs:
        \[
        \widehat{\mathrm{Var}}(\hat{\psi}_{\mathcal{V}}) = \frac{1}{n_{\mathcal{V}}^2}\sum_{i \in \mathcal{V}}\bigl(\hat{\phi}_{ik}(O_i)\bigr)^2 \;+\; \frac{1}{n_{\mathcal{V}^c}^2}\sum_{i \in \mathcal{V}^c}\bigl(\hat{\phi}_{ik}(O_i)\bigr)^2.
        \]

        \item Compute the t-statistic:
        \[
        T_{\mathcal{V}} = \frac{\hat{\psi}_{\mathcal{V}}}{\sqrt{\widehat{\mathrm{Var}}(\hat{\psi}_{\mathcal{V}})}}.
        \]

        \item Record the split that maximizes \(|T_{\mathcal{V}}|\), subject to minimum sample size constraints to ensure valid asymptotic approximations.
    \end{enumerate}

    \item \textbf{Recursive Partitioning:}  
    Recursively apply the above procedure to the resulting subpopulations, increasing the depth of the partitioning tree until a predefined maximum depth is reached or no further statistically significant splits are found.

    \item \textbf{Selection of Optimal Region:}  
    Among all considered splits, select the covariate \( W_j \) and split point \( s \) that yield the maximal statistically significant effect modification.
\end{enumerate}

\subsubsection*{Algorithm Implementation}

The detailed pseudocode in \textbf{Algorithm~\ref{alg:recursive_partitioning}} outlines the recursive partitioning approach, leveraging the t-statistic to identify the covariate regions that maximize the difference in subpopulation intervention effects.

\begin{algorithm}
\caption{Recursive Covariate Partitioning Based on T-Statistic}
\label{alg:recursive_partitioning}
\begin{algorithmic}[1]
\State \textbf{Input:} Data \( D \), covariates \( W \), IIEs \( \hat{\Delta} \), ICEs \( \hat{\phi} \), maximum depth \( d_{\text{max}} \), minimum observations \( \text{min\_obs} \)
\State \textbf{Output:} Rule defining the covariate region \( \mathcal{V} \) with maximal effect modification

\Procedure{T-Part}{$D$, $W$, $\hat{\Delta}$, $\hat{\phi}$, depth=0}
    \If{depth $= d_{\text{max}}$}
        \State \textbf{return} \Comment{Maximum depth reached}
    \EndIf
    \State Initialize \( \text{best\_split} \gets \text{Null} \), \( T_{\text{best}} \gets 0 \)
    
    \For{each covariate \( W_j \)}
        \For{each possible split point \( s \) in \( W_j \)}
            \State Define \( \mathcal{V} = \{ i : W_{ij} \leq s \} \), \( \mathcal{V}^c = \{ i : W_{ij} > s \} \)
            \If{ \( n_{\mathcal{V}} < \text{min\_obs} \) or \( n_{\mathcal{V}^c} < \text{min\_obs} \) }
                \State \textbf{continue}
            \EndIf
            \State Compute \( \hat{\psi}_{\mathcal{V}} \) and \( \widehat{\mathrm{Var}}(\hat{\psi}_{\mathcal{V}}) \) as above
            \State Compute \( T_{\mathcal{V}} = \dfrac{\hat{\psi}_{\mathcal{V}}}{\sqrt{ \widehat{\mathrm{Var}}(\hat{\psi}_{\mathcal{V}}) }} \)
            \If{ \( |T_{\mathcal{V}}| > |T_{\text{best}}| \) }
                \State \( T_{\text{best}} \gets T_{\mathcal{V}} \)
                \State \( \text{best\_split} \gets (W_j, s) \)
            \EndIf
        \EndFor
    \EndFor
    
    \If{\( \text{best\_split} \neq \text{Null} \) and \( |T_{\text{best}}| \) is statistically significant}
        \State Partition data based on \( \text{best\_split} \) into \( D_{\mathcal{V}} \) and \( D_{\mathcal{V}^c} \)
        \State Recursively apply \Call{T-Part}{$D_{\mathcal{V}}, W, \hat{\Delta}, \hat{\phi}, \text{depth} + 1$}
        \State Recursively apply \Call{T-Part}{$D_{\mathcal{V}^c}, W, \hat{\Delta}, \hat{\phi}, \text{depth} + 1$}
    \Else
        \State \textbf{return} \Comment{No significant split found}
    \EndIf
\EndProcedure

\State \textbf{Main Loop:}

\For{each exposure \( A_i \) in \( \boldsymbol{A} \)}
    \State Initiate \Call{T-Part}{$D$, $W$, $\hat{\Delta}_i$, $\hat{\phi}_i$, 0}
\EndFor

\State Select the covariate region \( \mathcal{V} \) with the maximal significant effect modification

\State \textbf{Return} \( \mathcal{V} \) and the corresponding effect modification estimate \( \hat{\psi}_{\mathcal{V}} \)
\end{algorithmic}
\end{algorithm}

\textbf{Remark.} By integrating the estimated influence functions into the partitioning algorithm, we ensure that the selection of covariate regions is guided not only by the magnitude of the estimated effect differences but also by their statistical significance. This approach helps to mitigate the risk of overfitting and the identification of spurious subpopulations due to random variability.

\subsection{Cross-Validation}

As discussed, sample splitting is required to avoid bias incurred by discovering subregions of the covariate space and estimating the effect of subpopulation intervention using the same data. To achieve this, we employ \( K \)-fold cross-validation, where the sample data are partitioned into \( K \) folds of approximately equal size. For example, consider \( K = 10 \). In each iteration \( k = 1, \ldots, K \):

\begin{enumerate}
    \item \textbf{Training and Validation Split:}
    \begin{itemize}
        \item The training set \( T_{n,k} \subset \{1, \ldots, n\} \) consists of the indices from all folds except the \( k \)-th fold, comprising \(\frac{K-1}{K}\) of the data (e.g., 90\% for \( K = 10 \)).
        \item The validation set \( V_{n,k} \subset \{1, \ldots, n\} \) consists of the indices in the \( k \)-th fold, comprising \(\frac{1}{K}\) of the data (e.g., 10\% for \( K = 10 \)).
    \end{itemize}
    
    \item \textbf{Stage 1: Subpopulation Identification and Parameter Estimation on Training Set:}
    \begin{itemize}
        \item Use the training data \( T_{n,k} \) to identify the maximal exposure-effect modifier sets within the mixture using the recursive partitioning algorithm described in Section \ref{sec:algorithm}.
        \item Estimate the nuisance parameters \( \hat{\overline{Q}} \) and \( \hat{g} \) using appropriate machine learning methods.
        \item Compute the individual intervention effects (IIE) and influence curve estimates (ICE) for each observation in \( T_{n,k} \).
    \end{itemize}
    
    \item \textbf{Stage 2: Parameter Estimation on Validation Set:}
    \begin{itemize}
        \item Apply the subpopulation rule \( \mathcal{V}_k \) identified from the training set to the validation data \( V_{n,k} \), assigning observations to \( \mathcal{V}_k \) or \( \mathcal{V}_k^c \).
        \item Estimate the subpopulation stochastic shift intervention parameter \( \Psi_{\mathcal{V}_k}(P) \) using the CV-TMLE approach with the nuisance parameters estimated from \( T_{n,k} \).
    \end{itemize}
\end{enumerate}

After completing all \( K \) folds, we aggregate the estimates across folds to obtain the final estimate of the effect modification parameter:
\begin{equation}
\hat{\Psi}_{\text{pooled}} = \frac{1}{K} \sum_{k=1}^K \hat{\Psi}_{\mathcal{V}_k}.
\end{equation}

If the identified subpopulation \( \mathcal{V}_k \) is consistent across folds, this pooled estimate leverages the full data for increased efficiency.

\subsection{Pooled TMLE}
\label{sec:pooled_tmle}

Upon completion of the cross-estimation procedure, a pooled TMLE update provides a summary measure of the oracle maximum effect modification parameter across \( K \)-folds. In each fold \( k \), we obtain estimates for \( \overline{Q} \) and \( g \) for the respective regions identified as the maximum effect-modifying regions and their complements. These regions may vary across different folds.

For the pooled TMLE update, we perform the following steps:
\begin{enumerate}
    \item \textbf{Stacking Nuisance Parameters:} Stack the estimates of the nuisance parameters \( \hat{\overline{Q}} \) and \( \hat{g} \) from each training sample \( T_{n,k} \).
    
    \item \textbf{Pooled TMLE Update:} Apply a pooled TMLE update on the cumulative initial estimates. This involves:
    \begin{itemize}
        \item Combining the initial outcome regression estimates and propensity score estimates across all folds.
        \item Performing a TMLE update on this pooled data to refine the estimates, ensuring they are both efficient and unbiased.
    \end{itemize}
    
    \item \textbf{Aggregating Estimates:} Compute the pooled estimate of the effect modification parameter by averaging the estimates across all folds:
    \begin{equation}
    \Psi_{\text{pooled}} = \frac{1}{K} \sum_{k=1}^K \Psi_{F_{T_{n,k}}}\left(V_{n,k}\right),
    \end{equation}
    where \( \Psi_{F_{T_{n,k}}} \) identifies the optimal exposure-covariate modifier set and formulates a plug-in estimator using the training data \( T_{n,k} \). \( V_{n,k} \) refers to the validation sample for deriving the estimates.
    
    \item \textbf{Interpretation of Pooled Estimates:} The pooled estimate \( \Psi_{\text{pooled}} \) represents an average across folds. If the identified subpopulations \( \mathcal{V}_k \) and \( \mathcal{V}_k^c \) are consistent across all folds (e.g., consistently identifying age \( < 22 \) for \( \mathcal{V} \) and age \( \geq 22 \) for \( \mathcal{V}^c \)), the pooled estimates are interpretable and provide increased statistical power. This leverages the full data, resulting in tighter confidence intervals and more reliable inference.
\end{enumerate}

\subsection{Pooled TMLE for Marginal Exposure Effects}
\label{sec:pooled_tmle_marginal}

When analyzing a mixture of multiple exposures, it is often important to assess the relative importance of each exposure component. Our approach facilitates this by estimating a stochastic shift for each exposure and deriving the corresponding influence curve estimates (ICE) to identify regions with maximal effect modification.

Since our target parameter requires stochastic shift interventions for each exposure as a preliminary step before recursive partitioning, we can efficiently leverage these estimates to obtain CV-TMLE for each exposure. The procedure is as follows:

\textbf{Utilizing Existing Estimates:} In the first stage of our methodology, we estimate stochastic shift interventions for each exposure \( A_i \) in the mixture \(\boldsymbol{A}\) and derive the corresponding ICEs. These estimates are obtained through the training phase of our cross-validation process and are applied to the validation folds.

\textbf{Pooled CV-TMLE Estimates:} After estimating the stochastic shifts and ICEs for each exposure across all folds, we aggregate these results to compute CV-TMLE estimates for each exposure shift. This pooling process involves:

\begin{itemize}
    \item Saving Validation Estimates: We store the TMLE-updated estimates of the outcome changes due to shifts in each exposure from the validation folds.
    
    \item Aggregating Estimates: By averaging these estimates across all folds, we obtain pooled CV-TMLE estimates for each exposure \( A_i \). This aggregation enhances the reliability and statistical power of our estimates.
\end{itemize}

\textbf{Variable Importance Measures:} The pooled CV-TMLE estimates for each exposure shift allow us to assess the relative importance of each exposure within the mixture. Specifically, these estimates quantify the impact of shifting each exposure \( \delta_i \) on the outcome, providing a clear measure of variable importance. This comprehensive estimation framework not only identifies regions with maximal effect modification but also offers insights into the marginal effects of each exposure, thereby supporting informed and targeted public health interventions.

In summary, because our target parameter inherently requires stochastic shift interventions for each exposure as part of the initial estimation process, we can directly utilize these results to obtain CV-TMLE estimates. This approach simplifies the workflow and enables the derivation of variable importance measures for the entire exposure mixture.

\subsection{Exposure Density Estimation}
\label{sec:density_estimation}

Accurate estimation of the conditional exposure densities \( g(A \mid W) \) is crucial for evaluating the effects of stochastic interventions within a causal inference framework. We employ two  approaches for density estimation: direct conditional density estimation and a reparameterization strategy that transforms density ratio estimation into a classification problem. Both methods are implemented using the Super Learner ensemble \cite{van2007super}, which optimally combines multiple algorithms to produce robust and flexible estimates.

\subsubsection{Direct Conditional Density Estimation}

The direct approach involves estimating the conditional density \( g(A \mid W) \) directly using a Super Learner composed of a diverse library of density estimation algorithms. This method leverages the strengths of various learners to capture complex relationships between exposures and covariates. However, direct density estimation can be challenging in high-dimensional or intricate settings due to the computational complexity and lack of models to estimate \( g(A \mid W) \) accurately.

\subsubsection{Reparameterization via Classification}

To address the limitations of direct density estimation, we implement a reparameterization strategy inspired by the work of \citet{diaz2023nonparametric} and on modified treatment policies (MTPs). This approach reframes the problem of estimating the density ratio \( \frac{g(A - \delta \mid W)}{g(A \mid W)} \) as a binary classification task. The procedure involves the following steps:

\begin{enumerate}
    \item \textbf{Augmented Dataset Construction:} Create an augmented dataset by duplicating the original data. In the duplicated subset, apply the stochastic shift intervention \( \delta \) to the exposures \( A \), while keeping the exposures in the original subset unchanged. Introduce a binary indicator variable \( Z \), where \( Z = 1 \) denotes shifted exposures and \( Z = 0 \) denotes unshifted exposures.
    
    \item \textbf{Classification Model Training:} Fit a Super Learner classifier to predict the probability \( P(Z = 1 \mid A, W) \) using the augmented dataset. This model estimates the likelihood that an observation has been subjected to the shift intervention.
    
    \item \textbf{Density Ratio Estimation:} Utilize Bayes' theorem to transform the predicted probabilities into the desired density ratios:
    \[
    \frac{g(A - \delta \mid W)}{g(A \mid W)} = \frac{P(Z = 1 \mid A - \delta, W)}{1 - P(Z = 1 \mid A, W)}.
    \]
\end{enumerate}

This reparameterization offers several key advantages:

\begin{itemize}
    \item \textbf{Computational Efficiency and Scalability:} By converting density ratio estimation into a classification problem, we can leverage highly optimized classification algorithms within the Super Learner framework, enhancing computational efficiency and scalability, especially in high-dimensional settings.
    
    \item \textbf{Numerical Stability:} Estimating density ratios through classification probabilities reduces the risk of extreme ratio estimates, thereby enhancing the numerical stability of our estimators.
    
    \item \textbf{Flexibility with Multiple Exposures:} This approach naturally extends to scenarios involving multiple exposures, facilitating future work to model the effect modification of joint exposure shifts.
\end{itemize}

Given these benefits, our simulations and empirical applications utilize the reparameterization strategy for density ratio estimation. This choice is motivated by its superior performance in terms of computational efficiency, estimator stability, which is particularly important in our case where the density ratio needs to be calculated for each exposure.

\section{Simulation Study}
\label{sec:simulation_study}
\subsection{Data Generation with Binary Modifiers}

To evaluate the performance of our proposed method, we conducted a simulation study using synthetic data that mirrors a realistic environmental epidemiology scenario. We generated a large population with 10000 observations to estimate ground truth of the subpopulation intervention effect. Three binary confounders, $W_1$, $W_2$, and $W_3$, were independently generated from Bernoulli distributions with a success probability of 0.5.

The exposures (\(A_1, A_2, A_3\)) were generated as follows:
\[
\begin{aligned}
A_1 &\sim \mathcal{N}(0.5 W_3 + 0.3 W_2 + 0.4 W_1, 1) \\
A_2 &\sim \mathcal{N}(0.3 W_2 + 0.3 W_1, 1) \\
A_3 &\sim \mathcal{N}(0.2 W_1, 1)
\end{aligned}
\]

The outcome \(Y\) was then simulated based on the exposures and covariates, incorporating effect modification effects:
\[
Y = 2 + 1 \cdot A_1 + 0.5 \cdot A_2 + 0.2 \cdot A_3 + 0.5 \cdot W_3 - 0.3 \cdot W_2 + 0.4 \cdot W_1 + 2 \cdot A_1 \cdot W_3 + \epsilon
\]
where \(\epsilon \sim \mathcal{N}(0, 1)\) represents random noise.

\subsection{Data Generation with Continuous Modifier}

To make detection of the modifying region more difficult and arguably more realistic compared to the binary case, we also create a DGP with a continuous covariate where the impact on the outcome jumps at a threshold. For this, we change $W_3$ to a continuous $\mathcal{N}(40, 10)$ and create a binary indicator for when $W_3 > 55$ and use this indicator in place of $W_3$ in the above outcome generation equation, but provide the \texttt{EffectXshift} algorithm the continuous covariate to determine whether it detects the correct threshold that leads to modification.

\subsection{Performance Metrics}
To establish the ground truth for the differential impacts of exposures, we applied a shift intervention to reduce \(A_1\) by 0.5 units, i.e., we set \(A_1 \leftarrow A_1 - 0.5\), and recalculated the outcome \(Y\) to obtain \(Y_{\delta}\). Other exposures remained unchanged. The difference between the shifted and unshifted outcomes (\(\Psi = Y_{\delta} - Y\)) was aggregated to estimate the true average effect within each level of the covariate \(W_3\), where the true effect modification was created. Here the true average in level 0 for $W_3$ is region $\mathcal{V}$ and level 1 is $\mathcal{V}^c$. 

We assessed the bias, variance, mean squared error (MSE), and coverage of our method across different sample sizes (\(n = 300, 500, 1000, 2000, 5000\)). For each sample size, we ran the simulation 100 times to evaluate the stability and reliability of our estimator. \textbf{Bias}: The average difference between the estimated and true effect. \textbf{Variance}: The variability of the estimator across simulations. \textbf{MSE}: Sum of the squared bias and variance, providing a comprehensive measure of estimator accuracy. \textbf{Coverage}: The proportion of times the true effect lies within the estimated 95\% confidence interval.

These metrics are essential for ensuring the \(\sqrt{n}\)-consistent behavior of our estimator, meaning that as the sample size increases, the estimator converges to the true parameter value at a rate proportional to the square root of the sample size. This property is crucial for an efficient estimator, as it guarantees precise and reliable estimates in large samples.

For our estimator to be asympotically unbaised we need to accurately detect the correct region with increasing sample size. As such we report metrics for \textbf{Accuracy, Precision, Recall and F1} statistics as well. 

\subsection{Simulation Process}

The simulation process involved the following steps: \textbf{1. Data Generation}: For each sample size, we generated datasets of the above data. \textbf{2. Estimator Application}: We applied our method to each dataset, to estimate the intervention effects within discovered subregions using 10-fold CV and default machine learning algorithms in the Super Learners (xgboost, random forest, glm) for the outcome regression and density estimation. \textbf{3. Metric Calculation}: For each sample size and iteration, we calculated the bias, variance, MSE, and coverage of the estimated effects. \textbf{4. Result Aggregation}: The results were aggregated to evaluate the overall performance of our method across different sample sizes.

\section{Simulation Results}
\label{sec:simulation_results}

\subsection{Identification of Maximum Effect Modifying Subregion}

For the binary effect modifier, for all iterations and sample sizes, we consistently identified the correct modifying variable, $W_3$ for the correct region, therefore accuracy, precision, recall and f1 were all 1.0. 

\begin{figure}[!h]
  \hspace*{0 cm}\includegraphics[scale = 0.3]{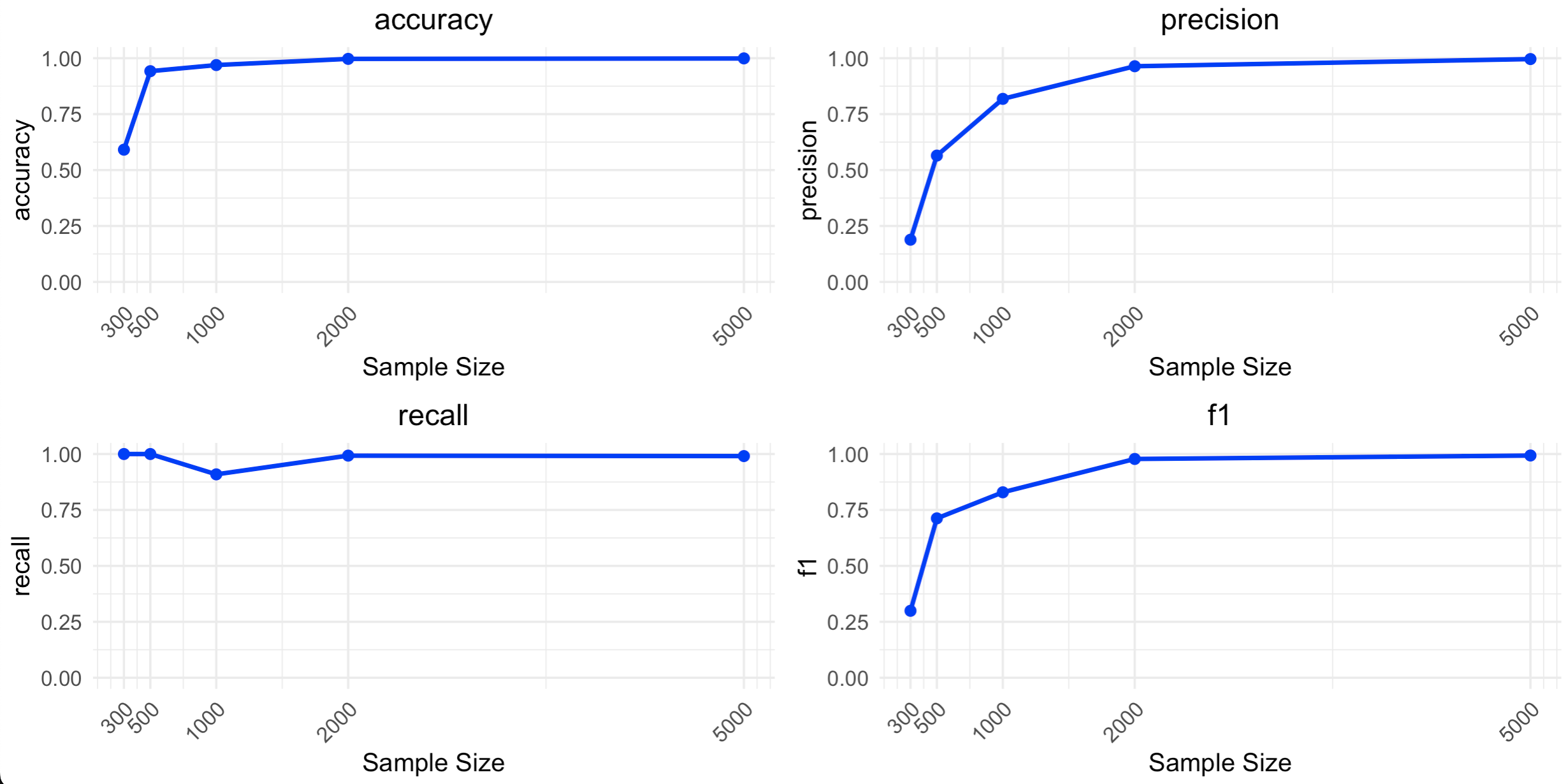}
  \caption{Subpopulation Detection Metrics for Continuous Modifier}
  \label{fig:cont_detection}
\end{figure}

The performance of our method in identifying the true subpopulation in continuous data improves as sample size increases. At a sample size of 300, the accuracy is 59.1\%, precision is 18.9\%, recall is 100\%, and the F1 score is 29.9\%. The high recall indicates perfect sensitivity, but the low precision suggests a high number of false positives. With a sample size of 500, accuracy improves to 94.2\%, precision to 56.5\%, recall remains at 100\%, and the F1 score rises to 71.2\%. The precision increase indicates fewer false positives. At 1000 samples, accuracy is 96.9\%, precision is 81.8\%, recall is 90.9\%, and the F1 score is 82.9\%. The balance between precision and recall reflects improved detection. For 2000 samples, accuracy is 99.7\%, precision is 96.4\%, recall is 99.3\%, and the F1 score is 97.8\%, showing minimal false positives and false negatives. At 5000 samples, the method nearly perfects performance with an accuracy of 99.9\%, precision of 99.6\%, recall of 99.1\%, and an F1 score of 99.3\%. \textbf{Figure \ref{fig:cont_detection}} shows these results.

\subsection{Binary Effect Modifier Simulation Results}

The results are summarized below based on bias, variance, mean squared error (MSE), and coverage probability.

\begin{itemize}
    \item \textbf{Bias and Variance:} The bias and variance generally decreased as the sample size increased. For smaller samples (e.g., $n = 300$), the bias was more pronounced in the measure in the $v^c$ region compared to the $v$ region, suggesting potential overfitting when sample sizes were limited. However, as the sample size increased, the bias significantly reduced, particularly evident in the $v^c$ region at $n = 5000$ where the bias was minimal ($-0.007$).

    \item \textbf{Mean Squared Error (MSE):} The MSE consistently decreased as the sample size increased, reflecting improvements in estimator accuracy. For instance, the MSE in the $v$ region decreased from 0.017 at $n = 300$ to 0.0001 at $n = 5000$. Similarly, in the $v^c$ region, the MSE decreased from 0.010 at $n = 300$ to 0.0001 at $n = 5000$.

    \item \textbf{Coverage Probability:} The coverage probability was stablazlied to 0.95 for the $v$ region across sample sizes but $v^c$ region maintained a coverage probability of 1, indicating overly conservative estimates which may be due to limited iterations or our DGP.
\end{itemize}

\subsection{Continuous Effect Modifier Simulation Results}

The results of the continuous effect modifier simulation are summarized below based on bias, variance, mean squared error (MSE), and coverage probability.

\begin{itemize}
    \item \textbf{Bias and Variance:} As the sample size increased, both bias and variance generally decreased. For the $v$ region, the bias reduced from 0.155 at $n = 300$ to a minimal 0.001 at $n = 5000$. Similarly, in the $v^c$ region, the bias decreased from 1.025 at $n = 300$ to 0.112 at $n = 5000$, indicating significant improvement in accuracy with larger sample sizes.

    \item \textbf{Mean Squared Error (MSE):} The MSE consistently decreased as the sample size increased, reflecting improvements in estimator accuracy. For the $v$ region, the MSE dropped from 0.026 at $n = 300$ to 0.0007 at $n = 5000$. In the $v^c$ region, the MSE decreased from 1.087 at $n = 300$ to 0.014 at $n = 5000$, demonstrating a substantial reduction in prediction error with larger samples.

    \item \textbf{Coverage Probability:} The coverage probability showed consistent improvement with increasing sample size. For the $v$ region, coverage increased from 0.80 at $n = 300$ to 0.95 at $n = 5000$, indicating better confidence interval reliability. In the $v^c$ region, coverage improved from 0.72 at $n = 300$ to 0.95 at $n = 5000$, suggesting that the true parameter value was increasingly captured within the calculated confidence intervals as sample size grew. Variability in coverage may be due to limited iterations in our simulation.
\end{itemize}

Overall, the results suggest that our estimator's mean squared error decreases at a rate proportional to $\sqrt{n}$, aligning with theoretical expectations. 

\begin{figure}[!h]
  \hspace*{0 cm}\includegraphics[scale = 0.3]{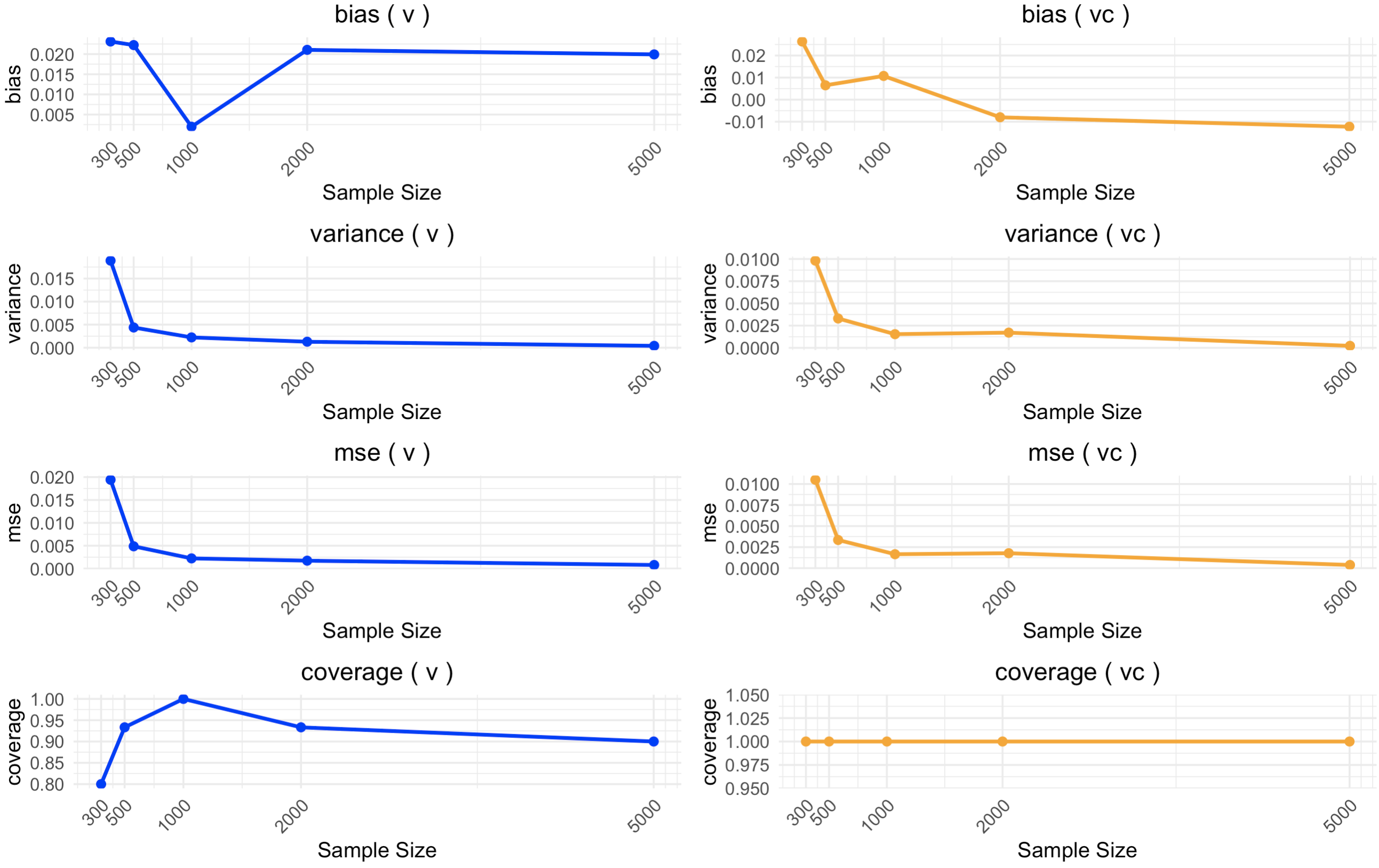}
  \caption{Simulation Results for Binary Effect Modifier}
  \label{fig:binary_sim_results}
\end{figure}

\begin{figure}[!h]
  \hspace*{0 cm}\includegraphics[scale = 0.3]{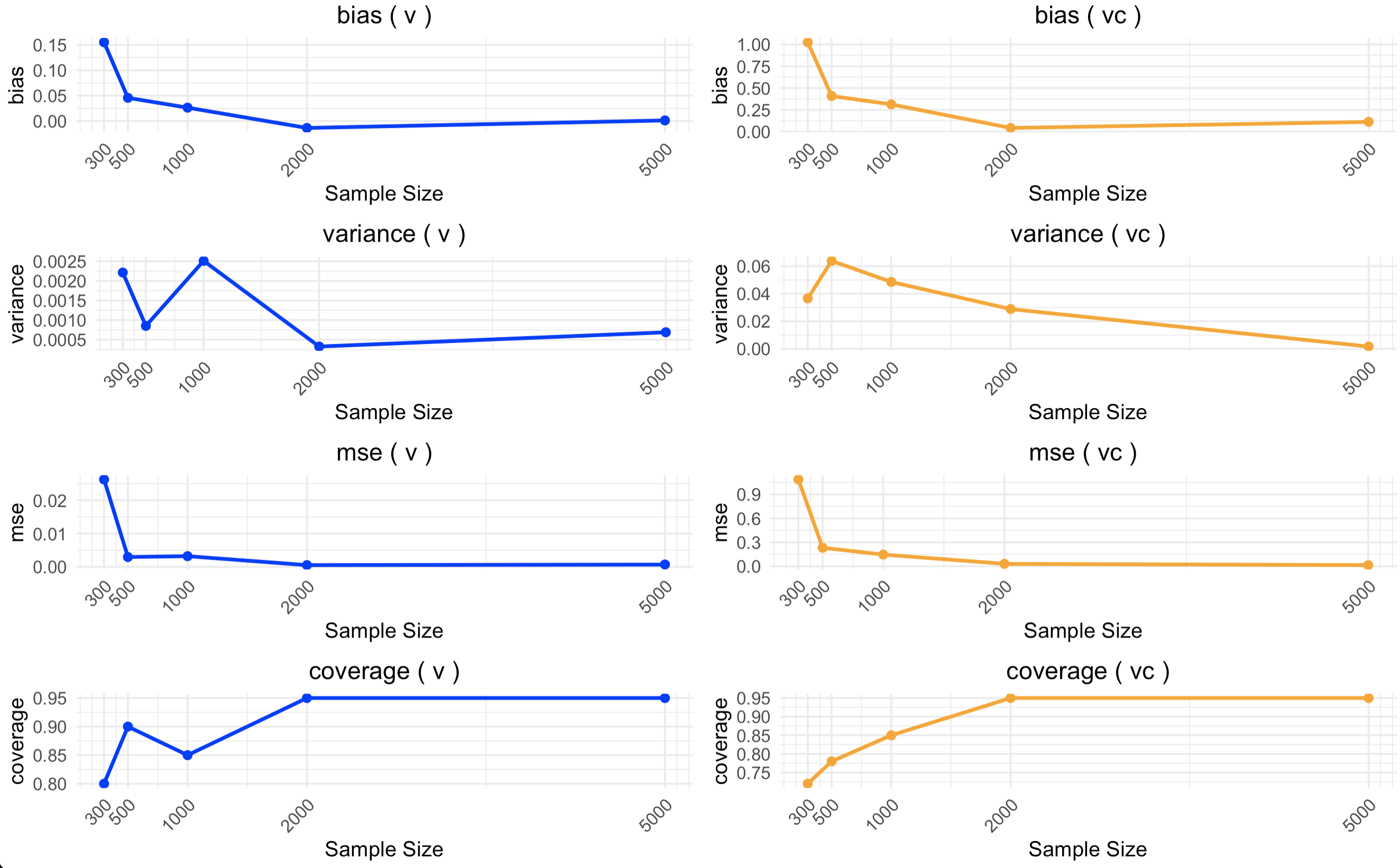}
  \caption{Simulation Results for Continuous Effect Modifier}
  \label{fig:cont_sim_results.png}
\end{figure}

\section{EffectXshift Applied to NHANES Dataset}
\label{sec:applied_nhanes}

The 2001-2002 National Health and Nutrition Examination Survey (NHANES) cycle provides a robust data set for our analysis, widely recognized for its credibility in the public health domain. This data set includes interviews with 11,039 individuals, of whom 4,260 provided blood samples and consented to DNA analysis. For our analysis, we focused on a subset of 1,007 participants, ensuring complete exposure data. This is comparable to the subset used by Mitro et al. \cite{mitro2016cross}, who investigated the association between persistent exposure to organic pollutants (POPs) binding to the aryl hydrocarbon receptor (AhR) and leukocyte telomere length (LTL). This data set was also used by Gibson et al., \cite{gibson2019} who applied various mixture methods to this data. For these reasons, this data set works nicely for comparison of the marginal variable importance effects given our stochastic shift interventions and provides evaluation of effect modification which has otherwise not been assessed for in these methods. 

Our study examined the effect of 18 congeners, including 8 non-dioxin-like PCBs, 2 non-ortho PCBs, 1 mono-ortho PCB, 4 dioxins, and 4 furans. All congeners were adjusted for lipid content in serum using an enzymatic summation method. Leukocyte telomere length was measured using quantitative polymerase chain reaction (qPCR) to determine the T/S ratio, which compares telomere length to a standardized reference DNA. 

Our analysis considered several covariates: age, sex, race/ethnicity, education level, BMI, serum cotinine levels, and blood cell distribution and count. Categories for race/ethnicity, education, and BMI were consistent with previous studies \cite{mitro2016cross, gibson2019}. Given evidence that telomere length can be influenced by factors such as smoking and age \cite{valdes_twin_2005, hoefnagel_review_2016}, our approach aimed to identify baseline covariate regions and specific exposures that exhibit the maximum differential impact due to a shift intervention. All covariate were included as potential effect modifiers. 

We utilized our \texttt{EffectXshift} package to perform this analysis. Using 10-fold cross-validation (CV), we explored the counterfactual changes in telomere length for a reduction of each exposure by 1 standard deviation. The standard deviation values for each exposure are as follows: PCB74 Lipid Adj (ng/g): -13589.77, 1,2,3,6,7,8-hxcdd Lipid Adj (pg/g): -40.36301, 1,2,3,4,6,7,8-hpcdd Lipid Adj (pg/g): -55.36449, 2,3,4,7,8-pncdf Lipid Adj (pg/g): -5.758755, 1,2,3,4,6,7,8-hxcdf Lipid Adj (pg/g): -10.67117, and 3,3',4,4',5-pcnb Lipid Adj (pg/g): -52.7907.

\subsection{NHANES Data Results}
\label{sec:nhanes_application}

We evaluated the impact of 16 persistent organic pollutants (POPs) on leukocyte telomere length (LTL). Our primary objective was to identify the baseline covariate region and the exposure with the maximum differential impact due to a reduction in exposure.

\subsection{Identification of Maximum Effect Modifiers}

The shift interventions were applied to a subset of uncorrelated exposures: PCB74, 1,2,3,6,7,8-hxcdd, 1,2,3,4,6,7,8-hpcdd, 2,3,4,7,8-pncdf Lipid Adj, 1,2,3,4,6,7,8-hxcdf, and 3,3',4,4',5-pcnb. Age was consistently identified as the maximum effect modifier across all analysis folds. The exposure 3,3',4,4',5-pcnb (LBXPCBLA) was found in 100\% of the folds, indicating a robust association. \textbf{Table \ref{tab:NHANES_EM}} shows the K-fold results. The change in outcome due to a 1 standard deviation reduction in 3,3',4,4',5-pcnb is shown under the column "Effect", inference is given in the standard error (SE) and confidence interval columns. "Modifier" is the region in the covariate space and its complement and Fold is the fold number. 

\begin{table}[htbp]
    \centering
    \caption{Effect Modification K-Fold Results}
    \begin{tabular}{lrrrrrlr}
        \hline
        \textbf{Exposure} & \textbf{Effect} & \textbf{SE} & \textbf{Lower CI} & \textbf{Upper CI} & \textbf{Modifier} & \textbf{Fold} \\
        \hline
        LBXPCBLA & 0.7047676 & 0.04008078 & 0.6262 & 0.7833 & age $\leq$ 14 & 1 \\
        LBXPCBLA & 0.5625930 & 0.08915217 & 0.3879 & 0.7373 & age $>$ 14 & 1 \\
        LBXPCBLA & 0.8017565 & 0.03205413 & 0.7389 & 0.8646 & age $\leq$ 11 & 2 \\
        LBXPCBLA & 0.7246825 & 0.08895869 & 0.5503 & 0.8990 & age $>$ 11 & 2 \\
        LBXPCBLA & 0.3222981 & 0.02627014 & 0.2708 & 0.3738 & age  $\leq$ 11 & 3 \\
        LBXPCBLA & 0.3271666 & 0.07242882 & 0.1852 & 0.4691 & age  $>$ 11 & 3 \\
        LBXPCBLA & -0.6316206 & 0.02034907 & -0.6715 & -0.5917 & age  $\leq$ 11 & 4 \\
        LBXPCBLA & -0.3153549 & 0.03551484 & -0.3850 & -0.2457 & age  $>$ 11 & 4 \\
        LBXPCBLA & 0.8372697 & 0.02145195 & 0.7952 & 0.8793 & age $\leq$ 14 & 5 \\
        LBXPCBLA & 0.6842051 & 0.07540448 & 0.5364 & 0.8320 & age  $>$ 14 & 5 \\
        LBXPCBLA & 0.7044444 & 0.02454143 & 0.6563 & 0.7525 & age  $\leq$ 11 & 6 \\
        LBXPCBLA & 0.6348496 & 0.06230732 & 0.5127 & 0.7570 & age  $>$ 11 & 6 \\
        LBXPCBLA & 0.7364454 & 0.02360700 & 0.6902 & 0.7827 & age  $\leq$ 17 & 7 \\
        LBXPCBLA & 0.6057689 & 0.08581898 & 0.4376 & 0.7740 & age  $>$ 17 & 7 \\
        LBXPCBLA & 0.6412914 & 0.02309030 & 0.5960 & 0.6865 & age  $\leq$ 11 & 8 \\
        LBXPCBLA & 0.5364844 & 0.06194912 & 0.4151 & 0.6579 & age  $>$ 11 & 8 \\
        LBXPCBLA & 0.7782908 & 0.01962983 & 0.7398 & 0.8168 & age  $\leq$ 11 & 9 \\
        LBXPCBLA & 0.5685745 & 0.07544209 & 0.4207 & 0.7164 & age  $>$ 11 & 9 \\
        LBXPCBLA & 0.2118344 & 0.01881629 & 0.1750 & 0.2487 & age  $\leq$ 17 & 10 \\
        LBXPCBLA & 0.3519853 & 0.08829586 & 0.1789 & 0.5250 & age  $>$ 17 & 10 \\
        \hline
    \end{tabular}
    \label{tab:NHANES_EM}
\end{table}

\subsection{Consistency of Results}

Age consistently emerged as the maximum effect modifier across all folds. The exposure 3,3',4,4',5-pcnb (LBXPCBLA) was identified in 100\% of the folds. Our findings indicate that a reduction in exposure to 3,3',4,4',5-pcnb by one standard deviation (52.7907) leads to an increase in LTL. This effect is more pronounced in younger populations, with central ages of 11, 14, and 17 consistently identified as split points. All these results were statistically significant based on the confidence intervals, underscoring the differential impact of this exposure across age.

\subsection{Pooled Oracle Estimates}

For each region and its complement, we provide the pooled oracle estimates. Here, $v$ denotes the region for the lower age group and $v^c$ for the higher age group. Below is \textbf{Table \ref{tab:oracle_NHANES}} of these pooled results.

\begin{table}[htbp]
    \centering
    \caption{Effect Modification Region Pooled Results}
    \begin{tabular}{lrrrrrr}
        \hline
        \textbf{Condition} & \textbf{Psi} & \textbf{Variance} & \textbf{SE} & \textbf{Lower CI} & \textbf{Upper CI} & \textbf{P-value} \\
        \hline
        \( v \)  & 0.4879 & 0.0001027 & 0.0101 & 0.5078 & 0.4680 & 0.0000 \\
        \( v^c \) & 0.3156 & 0.0010592 & 0.0325 & 0.2518 & 0.3794 & 3.11e-22 \\
        \hline
    \end{tabular}
    \label{tab:oracle_NHANES}
\end{table}

These estimates indicate that shifting 3,3',4,4',5-pcnb by one standard deviation (approximately -52.8) leads to an estimated increase in telomere length of about 0.488 in the lower age group ($v$) and about 0.316 in the higher age group ($v^c$).

To formally assess whether the difference between these two subgroups is statistically significant, we can compare the estimates and their variances directly. Letting \(\hat{\psi}_{\mathcal{V}} = \hat{\Psi}_v - \hat{\Psi}_{v^c} = 0.4879 - 0.3156 = 0.1723\), and assuming negligible covariance between the estimates, the variance of the difference is \(\widehat{\mathrm{Var}}(\hat{\psi}_{\mathcal{V}}) = 0.0001027 + 0.0010592 = 0.0011619\). Thus, \(\text{SE}(\hat{\psi}_{\mathcal{V}}) = \sqrt{0.0011619} \approx 0.0341\).

The corresponding z-statistic is:
\[
Z = \frac{0.1723}{0.0341} \approx 5.06,
\]
which yields a p-value well below 0.001. This confirms that the difference in the effect estimates between the two subpopulations is highly statistically significant.

\section{Conclusion}
\label{sec:conclusion}

Our study presents a novel methodology, implemented in the open-source package, \texttt{EffectXshift}, found on github, to identify subpopulations that are differentially affected by exposures and to estimate the effects of interventions within these subpopulations. Methodologically, our approach addresses several limitations of traditional causal inference methods by providing a data-driven, assumption-lean framework that leverages machine learning for both the estimation of exposure effects and the identification of effect modifiers. The use of stochastic shift interventions within our framework allows for more realistic and flexible modeling of exposure distributions, which is particularly relevant in environmental health studies, where exposures are often not binary or fixed. 

By first establishing an oracle target parameter of interest as a definition of effect modification and then employing a data-adaptive target parameter strategy, we are able to deliver interpretable results grounded in theory. This approach is fundamentally different from other metalearning approaches in estimating heterogeneous treatment effects. 

Applying this approach to the NHANES dataset, we demonstrated its ability to uncover significant effect modifiers, specifically highlighting age as a consistent modifier for 3,3',4,4',5-pcnb exposure and its association with leukocyte telomere length.

The findings of our analysis may have important implications for environmental health research and policy making. Identifying age as a significant effect modifier, our results suggest that younger individuals are more susceptible to the adverse effects of  3,3',4,4',5-pcnb exposure on telomere length. This information can inform targeted public health interventions aimed at reducing exposure to persistent organic pollutants (POPs) among vulnerable subpopulations, particularly younger individuals.

Despite the strengths of our approach, there are some limitations. The accuracy of our estimates depends on the correct estimation of the clever covariate (the ratio of densities). Estimating this ratio can be challenging, which is why we offer an alternative approach via classification. However, more research is necessary to ensure this reparameterization of the density estimation is sound in practice, although we show very little difference in results; our DGP is relatively simple. Additionally, while our method can handle complex and high-dimensional data, the computational cost may be significant for very large datasets. Future research should focus on further optimizing the computational efficiency of the \texttt{EffectXshift} package and exploring its applicability to other types of environmental exposures and outcomes.

In conclusion, the \texttt{EffectXshift} package provides a powerful tool for environmental health researchers to identify and estimate the differential impacts of exposures between subpopulations. By facilitating the discovery of vulnerable groups and informing targeted interventions, this methodology contributes to the broader goal of reducing health disparities and improving public health outcomes.

\bibliographystyle{unsrtnat}
\bibliography{references}

\begin{thebibliography}{21}
\providecommand{\natexlab}[1]{#1}
\providecommand{\url}[1]{\texttt{#1}}
\expandafter\ifx\csname urlstyle\endcsname\relax
  \providecommand{\doi}[1]{doi: #1}\else
  \providecommand{\doi}{doi: \begingroup \urlstyle{rm}\Url}\fi

\bibitem[Balbus and Malina(2009)]{balbus_identifying_2009}
John~M Balbus and Catherine Malina.
\newblock Identifying vulnerable subpopulations for climate change health effects in the united states.
\newblock \emph{Journal of Occupational and Environmental Medicine}, 51\penalty0 (1):\penalty0 33--37, 2009.
\newblock \doi{10.1097/JOM.0b013e318193e12e}.

\bibitem[Ruiz et~al.(2018)Ruiz, Becerra, Jagai, Ard, and Sargis]{ruiz_disparities_2018}
Daniel Ruiz, Marlene Becerra, Jyotsna~S Jagai, Kerry Ard, and Robert~M Sargis.
\newblock Disparities in environmental exposures to endocrine-disrupting chemicals and diabetes risk in vulnerable populations.
\newblock \emph{Diabetes Care}, 41\penalty0 (1):\penalty0 193--205, 2018.
\newblock \doi{10.2337/dc16-2765}.

\bibitem[Zorzetto et~al.(2024{\natexlab{a}})Zorzetto, Coker, Pearson, and Houghton]{zorzetto_confounder-dependent_2024}
Renata Zorzetto, Eric~S Coker, James~F Pearson, and Ceylan Houghton.
\newblock Confounder-dependent bayesian mixture modeling: Characterizing heterogeneity in causal effects of air pollution.
\newblock \emph{Environmental Health Perspectives}, 132\penalty0 (1):\penalty0 174--183, 2024{\natexlab{a}}.

\bibitem[Makri and Stilianakis(2008)]{MAKRI2008326}
Anna Makri and Nikolaos~I. Stilianakis.
\newblock Vulnerability to air pollution health effects.
\newblock \emph{International Journal of Hygiene and Environmental Health}, 211\penalty0 (3):\penalty0 326--336, 2008.
\newblock ISSN 1438-4639.
\newblock \doi{https://doi.org/10.1016/j.ijheh.2007.06.005}.
\newblock URL \url{https://www.sciencedirect.com/science/article/pii/S1438463907000971}.

\bibitem[Rothman et~al.(2008)Rothman, Greenland, and Lash]{Rothman2008}
Kenneth~J. Rothman, Sander Greenland, and Timothy~L. Lash.
\newblock \emph{Modern Epidemiology}.
\newblock Lippincott Williams \& Wilkins, Philadelphia, PA, 3 edition, 2008.
\newblock ISBN 9780781755641.

\bibitem[Samoli et~al.(2011)Samoli, Nastos, Paliatsos, Katsouyanni, and Priftis]{Samoli2011}
E~Samoli, PT~Nastos, AG~Paliatsos, K~Katsouyanni, and KN~Priftis.
\newblock Acute effects of air pollution on pediatric asthma exacerbation: Evidence of association and effect modification.
\newblock \emph{Environmental Research}, 111\penalty0 (3):\penalty0 418--424, 2011.

\bibitem[Analitis et~al.(2014)Analitis, Michelozzi, D'Ippoliti, de'Donato, Menne, Matthies, Atkinson, Iñiguez, Basagaña, Schneider, Lefranc, Paldy, Bisanti, and Katsouyanni]{Analitis2014}
Antonis Analitis, Paola Michelozzi, Daniela D'Ippoliti, Francesca de'Donato, Bettina Menne, Franziska Matthies, Richard~W. Atkinson, Carmen Iñiguez, Xavier Basagaña, Alexandra Schneider, Agnès Lefranc, Anna Paldy, Luigi Bisanti, and Klea Katsouyanni.
\newblock Effects of heat waves on mortality: Effect modification and confounding by air pollutants.
\newblock \emph{Epidemiology}, 25\penalty0 (1):\penalty0 15--22, 2014.
\newblock \doi{10.1097/EDE.0b013e31828ac01b}.

\bibitem[Gibson et~al.(2019)Gibson, Nunez, Abuawad, Zota, Renzetti, Devick, Gennings, Goldsmith, Coull, and Kioumourtzoglou]{gibson2019}
Elizabeth~A. Gibson, Yanelli Nunez, Ahlam Abuawad, Ami~R. Zota, Stefano Renzetti, Katrina~L. Devick, Chris Gennings, Jeff Goldsmith, Brent~A. Coull, and Marianthi-Anna Kioumourtzoglou.
\newblock An overview of methods to address distinct research questions on environmental mixtures: an application to persistent organic pollutants and leukocyte telomere length.
\newblock \emph{Environmental Health}, 18\penalty0 (1):\penalty0 76, 2019.

\bibitem[Coker et~al.(2018)Coker, Liverani, Su, and Molitor]{coker_multi-pollutant_2018}
Eric Coker, Silvia Liverani, Jason~G Su, and John Molitor.
\newblock Multi-pollutant modeling through examination of susceptible subpopulations using profile regression.
\newblock \emph{Current Environmental Health Reports}, 5\penalty0 (1):\penalty0 59--69, 2018.
\newblock \doi{10.1007/s40572-018-0181-2}.

\bibitem[Zorzetto et~al.(2024{\natexlab{b}})Zorzetto, Bargagli-Stoffi, Canale, and Dominici]{Zorzetto2024}
Dafne Zorzetto, Falco~J Bargagli-Stoffi, Antonio Canale, and Francesca Dominici.
\newblock Confounder-dependent bayesian mixture model: Characterizing heterogeneity of causal effects in air pollution epidemiology.
\newblock \emph{Biometrics}, 80\penalty0 (2):\penalty0 ujae025, 2024{\natexlab{b}}.
\newblock \doi{10.1093/biomtc/ujae025}.

\bibitem[Künzel et~al.(2019)Künzel, Sekhon, Bickel, and Yu]{kunzel2019metalearners}
Sören~R Künzel, Jasjeet~S Sekhon, Peter~J Bickel, and Bin Yu.
\newblock Metalearners for estimating heterogeneous treatment effects using machine learning.
\newblock \emph{Proceedings of the National Academy of Sciences}, 116\penalty0 (10):\penalty0 4156--4165, 2019.

\bibitem[Muñoz and van~der Laan(2012)]{DiazMunoz2012}
Iván~Díaz Muñoz and Mark van~der Laan.
\newblock Population intervention causal effects based on stochastic interventions.
\newblock \emph{Biometrics}, 68\penalty0 (2):\penalty0 541--549, June 2012.
\newblock \doi{10.1111/j.1541-0420.2011.01685.x}.
\newblock URL \url{https://www.ncbi.nlm.nih.gov/pmc/articles/PMC4117410/}.

\bibitem[Kennedy(2019)]{Kennedy2019}
Edward~H. Kennedy.
\newblock Nonparametric causal effects based on incremental propensity score interventions.
\newblock \emph{Journal of the American Statistical Association}, 114\penalty0 (526):\penalty0 645--656, 2019.
\newblock \doi{10.1080/01621459.2018.1425625}.
\newblock URL \url{https://doi.org/10.1080/01621459.2018.1425625}.

\bibitem[van~der Laan and Rose(2011)]{vanderLaan2011}
Mark~J. van~der Laan and Sherri Rose.
\newblock \emph{Targeted Learning: Causal Inference for Observational and Experimental Data}.
\newblock Springer, New York, NY, 2011.
\newblock ISBN 9781441997821.
\newblock \doi{10.1007/978-1-4419-9782-1}.
\newblock URL \url{https://doi.org/10.1007/978-1-4419-9782-1}.

\bibitem[Jain(2015)]{Jain2016}
Richa~B. Jain.
\newblock Contribution of firefighting to exposures of poly- and perfluoroalkyl substances (pfas) in women firefighters from san francisco, california.
\newblock \emph{Environmental Research}, 142:\penalty0 661--670, 2015.
\newblock \doi{10.1016/j.envres.2015.07.002}.
\newblock URL \url{https://doi.org/10.1016/j.envres.2015.07.002}.

\bibitem[Li et~al.(2022)Li, Rosete, Coyle, Phillips, Hejazi, Malenica, Arnold, Benjamin-Chung, Mertens, and Jr]{Li2022}
Haodong Li, Sonali Rosete, Jeremy Coyle, Rachael~V. Phillips, Nima~S. Hejazi, Ivana Malenica, Benjamin~F. Arnold, Jade Benjamin-Chung, Andrew Mertens, and John M.~Colford Jr.
\newblock Evaluating the robustness of targeted maximum likelihood estimators via realistic simulations in nutrition intervention trials.
\newblock \emph{Statistics in Medicine}, 41\penalty0 (3):\penalty0 417--433, 2022.
\newblock \doi{10.1002/sim.9348}.
\newblock URL \url{https://doi.org/10.1002/sim.9348}.
\newblock Funding information: Bill and Melinda Gates Foundation, OPP1165144.

\bibitem[van~der Laan et~al.(2007)van~der Laan, Polley, and Hubbard]{van2007super}
Mark~J van~der Laan, Eric~C Polley, and Andrew~E Hubbard.
\newblock Super learner.
\newblock \emph{Statistical applications in genetics and molecular biology}, 6\penalty0 (1):\penalty0 Article 11, 2007.
\newblock \doi{10.22002/APGM.30714}.
\newblock URL \url{https://doi.org/10.22002/APGM.30714}.

\bibitem[D{\'\i}az et~al.(2023)D{\'\i}az, Williams, Hoffman, and Schenck]{diaz2023nonparametric}
Iv{\'a}n D{\'\i}az, Nicholas Williams, Katherine~L Hoffman, and Edward~J Schenck.
\newblock Nonparametric causal effects based on longitudinal modified treatment policies.
\newblock \emph{Journal of the American Statistical Association}, 118\penalty0 (542):\penalty0 846--857, 2023.
\newblock \doi{10.1080/01621459.2021.1955691}.
\newblock URL \url{https://doi.org/10.1080/01621459.2021.1955691}.

\bibitem[Mitro et~al.(2016)Mitro, Birnbaum, Needham, and Zota]{mitro2016cross}
Susanna~D. Mitro, Linda~S. Birnbaum, Larry~L. Needham, and Ami~R. Zota.
\newblock Cross-sectional associations between exposure to persistent organic pollutants and leukocyte telomere length among u.s. adults in nhanes, 2001-2002.
\newblock \emph{Environmental Health Perspectives}, 124\penalty0 (5):\penalty0 651--658, 2016.

\bibitem[Valdes et~al.(2005)Valdes, Andrew, Gardner, Kimura, Oelsner, Cherkas, Aviv, and Spector]{valdes_twin_2005}
Ana~M. Valdes, Toby Andrew, John~P. Gardner, Masayuki Kimura, Elizabeth Oelsner, Lynn~F. Cherkas, Abraham Aviv, and Tim~D. Spector.
\newblock Obesity, cigarette smoking, and telomere length in women.
\newblock \emph{The Lancet}, 366\penalty0 (9486):\penalty0 662--664, 2005.

\bibitem[Hoefnagel et~al.(2016)Hoefnagel, Janssen, Jaspers, de~Haan, and Ottenheijm]{hoefnagel_review_2016}
Sander J.~M. Hoefnagel, Jolanda A. M. J.~L. Janssen, Ronald~T. Jaspers, Albert de~Haan, and Coen A.~C. Ottenheijm.
\newblock The influence of lifestyle factors on telomere length in the adult population.
\newblock \emph{Journal of Gerontology}, 71\penalty0 (12):\penalty0 1467--1474, 2016.

\end{thebibliography}

\end{document}